\documentclass[12pt]{iopart}
  \expandafter\let\csname equation*\endcsname\relax
  \expandafter\let\csname endequation*\endcsname\relax
\usepackage{amsmath}
\usepackage{graphicx,eucal}
\usepackage{graphicx}
\usepackage{float}
\usepackage{cite}
\usepackage{subfigure}
\begin{document}

\title{Choice-Driven Phase Transition in Complex Networks}

\author{P.~L.~Krapivsky} \address{Department of Physics, Boston
  University, Boston, MA 02215}

\author{S.~Redner} \address{Department of Physics, Boston University,
  Boston, MA 02215 and Santa Fe Institute, 1399 Hyde Park Road, Santa Fe,
  New Mexico 87501, USA}

\begin{abstract}
  We investigate \emph{choice-driven} network growth.  In this model, nodes
  are added one by one according to the following procedure: for each
  addition event a set of target nodes is selected, each according to linear
  preferential attachment, and a new node attaches to the target with the
  highest degree.  Depending on precise details of the attachment rule, the
  resulting networks has three possible outcomes: (i) a non-universal
  power-law degree distribution; (ii) a single macroscopic hub (a node whose
  degree is of the order of $N$, the number of network nodes), while the
  remainder of the nodes comprises a non-universal power-law degree
  distribution; (iii) a degree distribution that decays as $(k\,\ln k)^{-2}$
  at the transition between cases (i) and (ii).  These properties are robust
  when attachment occurs to the highest-degree node from at least two
  targets.  When attachment is made to a target whose degree is \emph{not}
  the highest, the degree distribution has the ultra-narrow
  double-exponential form $\exp(-{\rm const.}\times e^k)$, from which the
  largest degree grows only as $\ln\ln N$.
\end{abstract}
\maketitle

\section{Introduction}

Choice plays an essential role in queuing and optimization theory
\cite{karp,broder,adler,michael,luczak}, in the structure of random recursive
trees \cite{RPC} and evolving random graphs \cite{BF,BK,SW}, in explosive
percolation \cite{ASS,Ziff09,FL09,KKK09,RF09,DM10,Dor10,RW11,Maya11}, and in
the control of avalanches in self-organized criticality~\cite{NBD13}.  We all
familiar with choice in grocery checkout, customs, and security lines, where
we would like to be in the line with the shortest waiting time.  Picking one
of $N$ lines at random results in a maximal waiting time of the order of $\ln
N$. If instead one initially selects two lines at random and then chooses the
line with the smaller number of customers, the maximal waiting time drops to
$O(\ln \ln N)$. Further increasing the number of initially selected lines
improves the maximal waiting time only by a constant factor, thereby
illustrating the ``power of two choices''
\cite{karp,broder,adler,michael,luczak}.

Growing networks with choice were investigated in \cite{RPC}, where the
choice was made to attach the new node to the node closest to the root.
Choice has also been implemented in evolving random graphs (networks with
fixed number of nodes and growing number of links), where it has been shown
that appropriate choice may delay~\cite{BF} or speed up~\cite{BK,SW} the
appearance of the giant component.  One particular example of choice-driven
link addition in evolving random graphs has recently attracted considerable
attention \cite{ASS,Ziff09,FL09,KKK09,RF09,DM10,Dor10,RW11,Maya11}, as it
leads a percolation transition which is explosive in character.

\begin{figure}[ht]
\centerline{\includegraphics[width=0.45\textwidth]{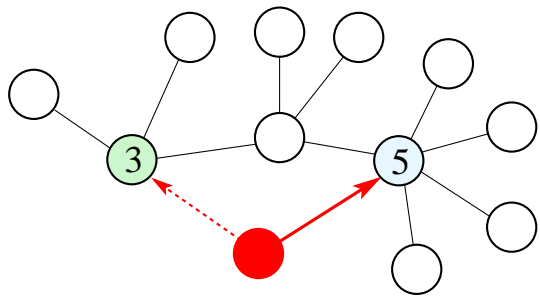}}
\caption{Illustration of network growth by greedy choice.  Two nodes (shaded)
  in the network are selected according to preferential attachment.  A new
  node (solid) attaches to the target with the larger degree, in this case,
  degree 5.}
\label{model}  
\end{figure}

In this work, we determine how a degree-based choice affects the growth of
complex networks~\cite{reviews}.  Instead of a new node attaching to a target
node according to a specified rate, we select a fixed number of targets
according to this rate and the new node attaches to the target with the
largest degree---``greedy'' choice (Fig.~\ref{model}).  When the targets are
selected randomly and independent of their degrees~\cite{RPC}, it was found
that the degree distribution decays exponentially with degree, but at a
slower rate than in the case with no choice.  When the targets are selected
according to the preferential attachment mechanism, the effect of the choice
is much more dramatic as we show below.

As an example, consider the situation where {\em two} targets are
provisionally selected, each with the probability proportional to
$A_k=k+\lambda$ for a target of degree $k$. Then our results can be
summarized as follows.  For $\lambda>0$, the network has a degree
distribution with an algebraic tail that possesses a non-universal exponent
(i.e., dependent on $\lambda$).  This exponent is smaller than in the case of
no choice; thus choice broadens the degree distribution.  For $\lambda=0$
(strictly linear preferential attachment), the degree distribution has a
power-law tail with the smallest possible exponent that is consistent with
the network remaining sparse.  More precisely, the fraction of nodes of
degree $k$ asymptotically decays as $(k\,\ln k)^{-2}$, with the logarithmic
factor ensuring that the network is sparse.  For $-1<\lambda<0$, a macrohub
(a node whose degree grows linearly with the number of nodes in the network)
emerges; the remainder of the degree distribution is still characterized by a
non-universal algebraic tail. These properties are qualitatively robust for
greedy choice with at least two alternatives, although the critical value of
$\lambda$ depends on the number of alternatives; in the case when $p$ target
nodes are provisionally selected, then $\lambda_c= p-2$.

\begin{figure}[ht]
\centerline{\subfigure[]{\includegraphics[width=0.5\textwidth]{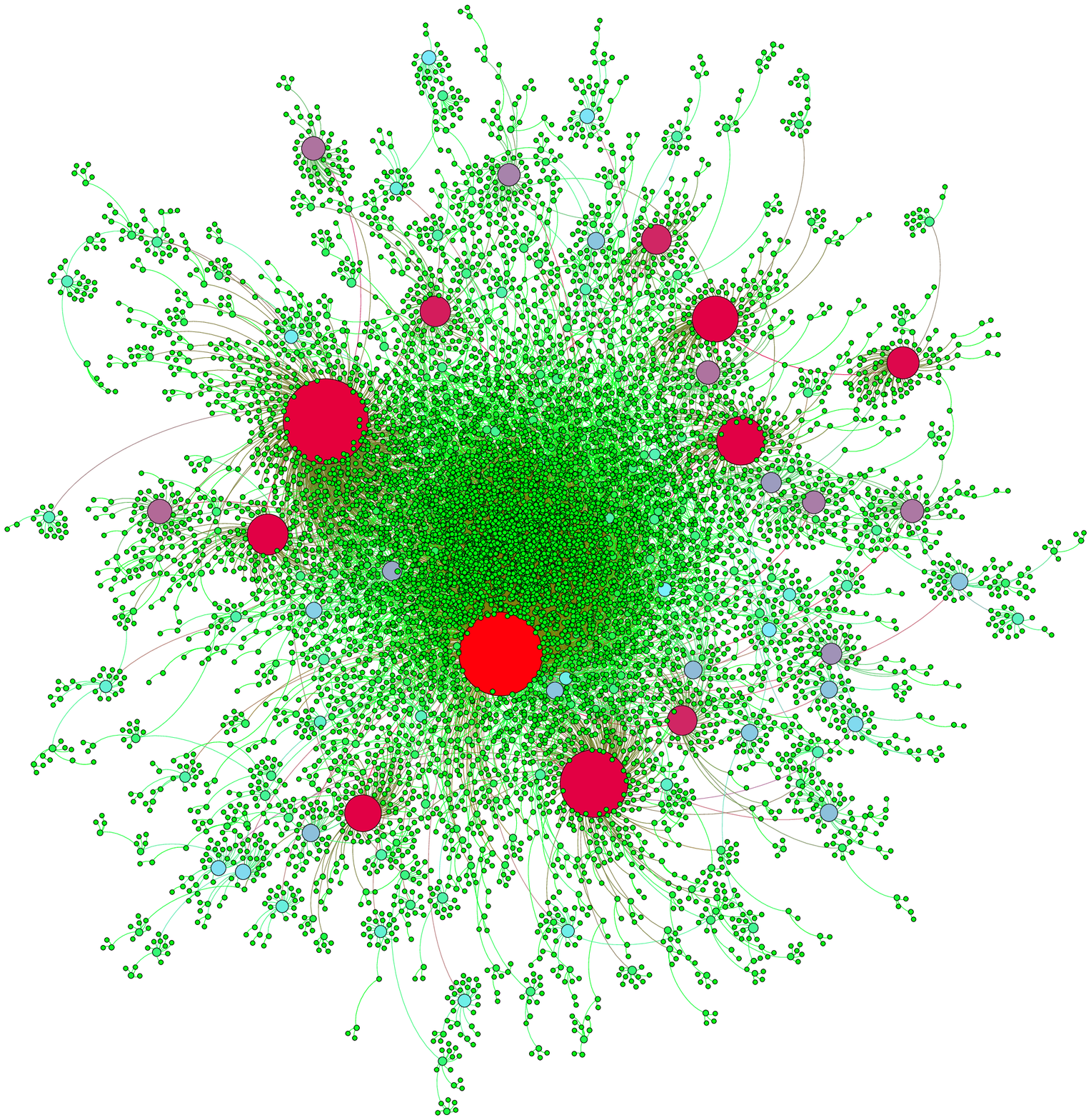}}
\subfigure[]{\includegraphics[width=0.5\textwidth]{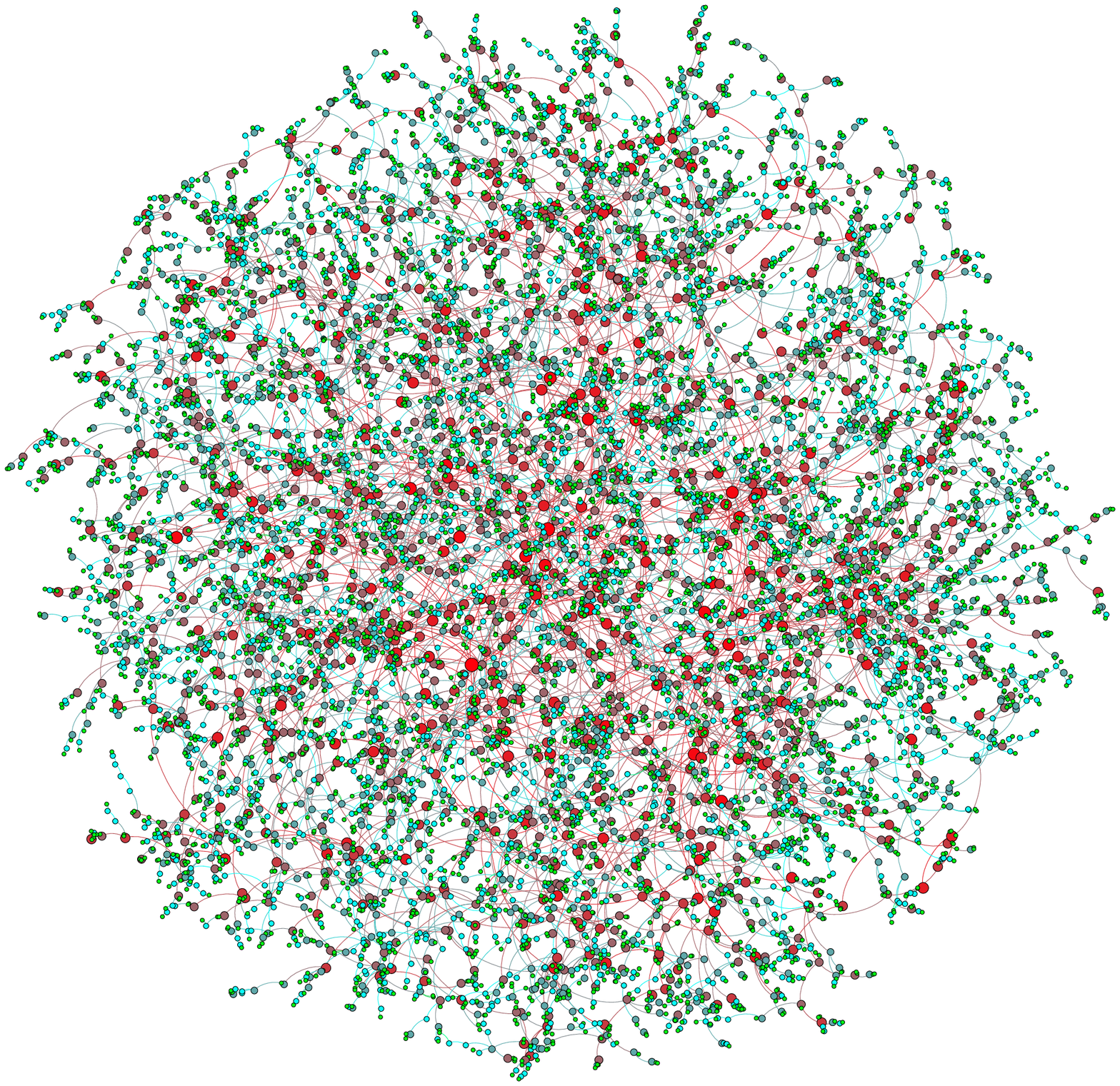}}}
\caption{Example networks of $10^4$ nodes that are grown by strictly linear
  preferential attachment for (a) greedy and (b) meek choice from two
  alternatives.  The maximal degree is 3399 in (a) and 8 in (b).  Red are
  high-degree nodes.}
\label{pix}  
\end{figure}

In contrast, when attachment occurs to a target whose degree is less than the
largest among the target set---which we term ``meek choice''---a
double-exponential degree distribution arises, where $n_k\sim \exp(-{\rm
  const.}\times e^k)$.  Somewhat surprisingly, this behavior occurs even if
attachment occurs to the second-largest out of a large number of targets.
Thus greedy choice is the unique case and all other less greedy attachment
choices lead to a double-exponential degree distribution.  Two examples of
small networks grown by greedy and meek choice from two alternatives are
shown in Fig.~\ref{pix}.

\section{Greedy Choice}

\subsection{Two Alternatives}

We start by studying the degree distribution in networks where growth is
driven by greedy choice between two alternatives.  Let $N_k(N)$ be the number
of nodes of degree $k$ when the network contains $N$ total nodes.  Although
the $N_k(N)$ are random variables, fluctuations in these quantities are small
when the network is large.  We thus focus on the averages $\langle
N_k(N)\rangle$ in the limit of large $N$, where we may replace
$N_k(N\!+\!1)-N_k(N)$ by $dN_k/dN$.  We also drop the angle brackets
henceforth.

The evolution of the degree distribution in this greedy choice model is
governed by the master equations
\begin{align}
\label{HD_long}
\frac{dN_k}{dN} &= \frac{A_{k-1}N_{k-1}}{A}
\sum_{j<k-1} \frac{A_{j}N_j}{A/2}
-\frac{A_kN_{k}}{A}\sum_{j<k} \frac{A_jN_j}{A/2} \nonumber\\
& +\left[\frac{A_{k-1}N_{k-1}}{A}\right]^2 
-\left[\frac{A_kN_{k}}{A}\right]^2+\delta_{k,1}\,.
\end{align}
Here $A_k$ is the rate at which a node of degree $k$ is selected as a
potential target and $A=\sum_jA_jN_j$ the total rate.  The first term on the
right-hand side of Eq.~\eqref{HD_long} accounts for the increase in $N_k$ due
to the new node attaching to a node of degree $k\!-\!1$.  Such an event
occurs if the two initial targets have degrees $k\!-\!1$ and $j<k\!-\!1$.
The complementary gain term has a similar origin, while the quadratic terms
on the second line account for events where the two targets have the same
degree.  The master equations satisfy the sum rules $\sum_{k\geq 1}N_k = N$
and $\sum_{k\geq 1}k N_k = 2(N\!-\!1)$.

In the following, we focus on the class of shifted linear attachment rates
given by $A_k=k+\lambda$.  In this case the total rate becomes
$A=\sum_jA_jN_j=(2+\lambda)N-2$.  We are interested in the $N\to\infty$
limit, so we simply write $A=\sum_jA_jN_j=(2+\lambda)N$.  The fraction of
nodes of fixed degree becomes size independent when $N\to\infty$, so that
$N_k(N) \to N n_k$ (see, e.g., \cite{GNR,book}).  Using this fact, we recast
\eqref{HD_long} into
\begin{align}
\label{HD_shift}
n_k = \frac{\psi_{k-1}-\psi_k}{(2+\lambda)^2/2}
\sum_{j<k} \psi_j
 -\frac{\psi_{k-1}^2+\psi_k^2}{(2+\lambda)^2}+\delta_{k,1}\,,
\end{align}
where $\psi_k\equiv(k\!+\!\lambda)n_k$

Let us now specialize to strictly linear preferential attachment, or
$\lambda=0$.  The solutions to the first few of the recurrences
\eqref{HD_shift} can be found straightforwardly and give
\begin{align}
\label{nigreedy}
n_1 &= 2\sqrt{2}-2 \approx 0.82843 \,, \nonumber \\
n_2 &= \frac{1}{2} - \sqrt{2} + \frac{1}{2} \sqrt{21-12\sqrt{2}}\approx  0.08945\,, \\
n_3 &=\frac{1}{9}  - \frac{1}{3}\sqrt{21-12\sqrt{2}}
+\frac{2}{9}\sqrt{70-6\sqrt{21-12\sqrt{2}} -36\sqrt{2}}\approx 0.03179\,,\nonumber
\end{align}
etc.  To obtain the asymptotic form of the degree distribution, it is
convenient to analyze \eqref{HD_shift} in the continuum approximation.  To
lowest order, we use the asymptotic behavior $\sum_{j<k} jn_j\to 2$ as
$k\to\infty$, which follows from $\sum_{k\geq 1}k N_k = 2(N\!-\!1)$, and we
also ignore the terms on the second line.  These approximations simplify
Eq.~\eqref{HD_shift} to $(kn_k)'=-n_k$, which gives $n_k\sim k^{-2}$.
However, this solution cannot be correct, as the sum $\sum_{k\geq 1}kn_k$
logarithmically diverges.  The inconsistency arises because the terms that
were dropped are of the same order, namely $k^{-2}$, as those in the
approximate equation $(kn_k)'=-n_k$.

\begin{figure}
\centerline{\includegraphics[width=0.65\textwidth]{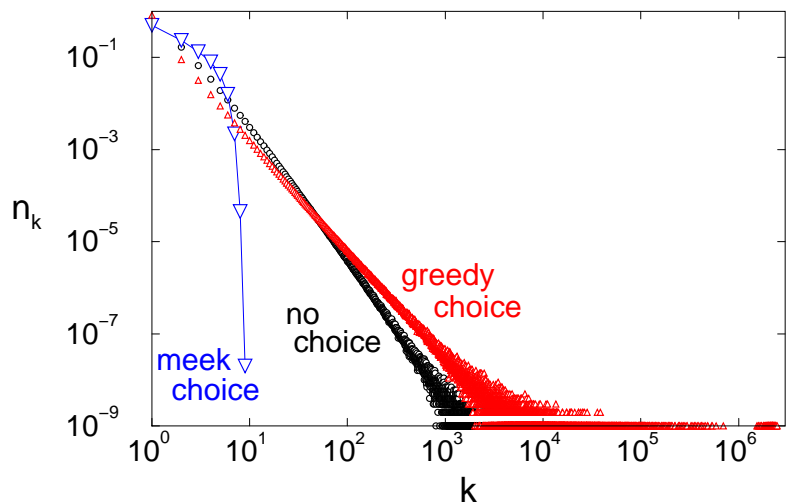}}
\caption{Influence of choice from two alternatives on the degree
  distributions of networks grown by strictly linear preferential attachment.
  The distribution without choice asymptotically decays as $k^{-3}$.  Data
  are based on $10^2$ realizations of $10^7$ nodes.}
\label{nks}  
\end{figure}

As will become plausible with hindsight, a logarithmic correction in the
asymptotic degree distribution can be anticipated.  We thus seek a solution
of the form
\begin{equation}
\label{u:def}
n_k=k^{-2} u(\ell), \qquad \ell = \ln k\,.
\end{equation}
Substituting this ansatz into \eqref{HD_shift}, keeping all terms, and using
the continuum approximation, gives
\begin{subequations}
\begin{equation}
\label{DE_long}
2u=\left(u-\frac{du}{d\ell}\right)\int_0^\ell dx u(x) - u^2\,,
\end{equation}
or, in terms of the cumulative variable $v(\ell) = \int_0^\ell dx\, u(x)$,
\begin{equation}
\label{DE_uv}
2=\left(1-\frac{du}{dv}\right)v - u\,,
\end{equation}
where we now view $u$ as a function of $v$.  This equation can be rewritten as
$(2-v)dv +udv+vdu=0$, with solution $2v-\tfrac{1}{2}v^2+uv=2$. (The
integration constant is set by the sum rule $\sum_{k\geq 1}kn_k=2$, which
implies $v(\infty)=2$ and $u(\infty)=0$.)~ Thus
\begin{equation}
u = \frac{dv}{d\ell} = \frac{2}{v}\left(1-\frac{v}{2}\right)^2 ~.
\end{equation}
Integrating gives 
\begin{equation}
\frac{\ell}{2} = \ln\left(1-\frac{v}{2}\right) + \frac{v}{2-v}~,
\end{equation}
\end{subequations}
or $2-v\simeq 4/\ell$, as $\ell\to\infty$.  Combining this result with
$v(\ell) = \int_0^\ell dx\, u(x)$ ultimately leads to $u\simeq 4/\ell^2$, so
that the asymptotic degree distribution is (see Fig.~\ref{nks})
\begin{equation}
\label{H_asymp}
n_k \simeq \frac{4}{k^2\,(\ln k)^2}~.
\end{equation}
Attempting a power-law fit to the data for $n_k$ versus $k$ leads to an
effective exponent that q appears to be slowly changing with $k$; this is
often the symptom of a logarithmic correction, as predicted by
\eqref{H_asymp}.

This slow decay of the degree distribution implies the existence of an almost
macroscopic hub---a node whose degree is nearly of the order of $N$.  To
estimate this maximal degree $k_\text{max}$ in a network that contains $N$
nodes, we apply the standard extremal criterion~\cite{G58} that there is of
the order of one node with degree $k_{\rm max}$ or larger,
\begin{equation}
\label{max}
\sum_{k\geq k_\text{max}}n_k \sim \frac{1}{N}~,
\end{equation}
to the degree distribution \eqref{H_asymp} to give
\begin{equation}
k_\text{max} \sim \frac{N}{(\ln N)^2}~;
\end{equation}
that is, a maximal degree that is almost of the order of $N$.

For shifted linear preferential attachment, $A_k=k+\lambda$, the degree
distribution without choice has the closed form~\cite{GNR}
\begin{equation}
  n_k =
  (2\!+\!\lambda)\,\frac{\Gamma(3\!+\!2\lambda)}{\Gamma(1\!+\!\lambda)}\,\,\,\frac{\Gamma(k\!+\!\lambda)}{\Gamma(k\!+\!3\!+\!2\lambda)}~, 
\end{equation}
whose asymptotic behavior is the non-universal power law $n_k\sim
k^{-(3+\lambda)}$.  (Note that $\lambda>-1$, so that attachment can occur to
nodes of degree 1.)

A convenient way to implement shifted linear preferential attachment is by
the redirection algorithm~\cite{GNR,book,kum}.  This algorithm consists of:
(i) selecting a target node uniformly at random from the existing network;
(ii) a new node either attaches to this target with probability $1-r$ or to
the parent of the target with probability $r$, where $r=(2+\lambda)^{-1}$.
This algorithm exactly reproduces network growth by shifted linear
preferential attachment with shift $\lambda$, where the redirection
probability is related to $\lambda$ via $r=(2+\lambda)^{-1}$.  This algorithm
is extremely simple and efficient, as the time to simulate a network of $N$
nodes scales linearly with $N$.

\begin{figure}
\centerline{\subfigure[]{\includegraphics[width=0.45\textwidth]{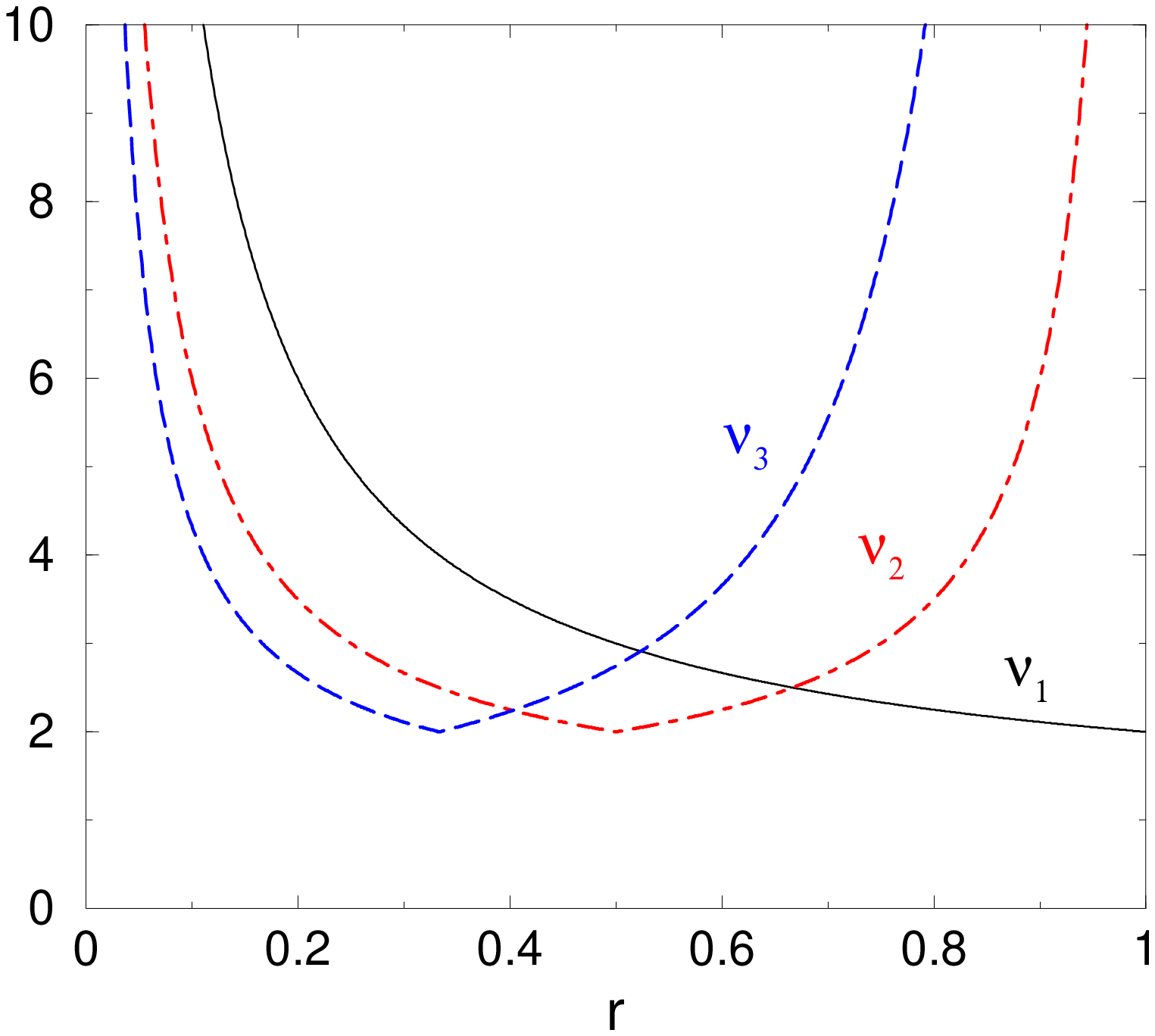}}\quad
\subfigure[]{\includegraphics[width=0.5\textwidth]{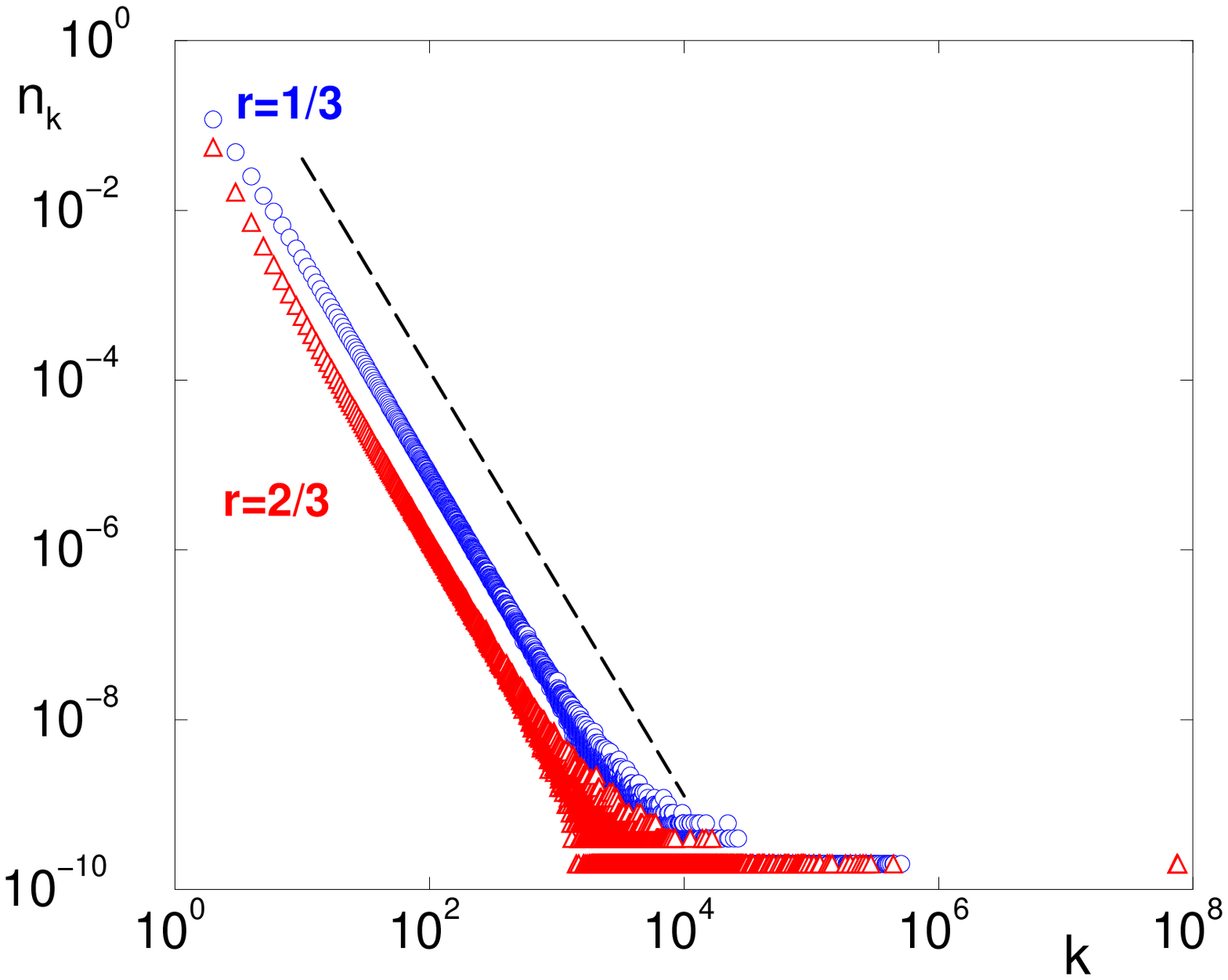}}}
\caption{(a) The exponents $\nu_1=2+\frac{1}{r}$, $\nu_2$ from
  Eq.~\eqref{nu2}, and $\nu_3$ from Eq.~\eqref{nu3}.  (b) Representative
  degree distributions for shifted linear preferential attachment with greedy
  choice for redirection probabilities $r=\frac{1}{3}$ and $r=\frac{2}{3}$
  for 50 realizations of a network of $10^8$ nodes.  The dashed line
  corresponds to exponent $\nu_2=2.5$, as given by \eqref{nu2} and
  \eqref{nu3}.  For $r=\frac{1}{3}$, the isolated data point at $k=7.5\times
  10^7$ corresponds to macrohubs whose degree is given by Eq.~\eqref{hub}.}
\label{nk}  
\end{figure}

We now determine how greedy choice affects the degree distribution when the
network grows by \emph{positive} shifted linear preferential attachment,
$A_k=k+\lambda$ with $\lambda >0$.  For large $k$, we again drop the
quadratic terms in \eqref{HD_shift}, replace $\sum_{j<k} (j\!+\!\lambda)n_j$
by $\sum_{j\geq 1} (j\!+\!\lambda)n_j = 2\!+\!\lambda$, and employ the
continuum approximation.  It may subsequently be verified that the dropped
terms are indeed subdominant when $\lambda>0$.  These steps yield
$(kn_k)'=-(2\!+\!\lambda)n_k/2$, with solution $n_k\sim k^{-(2+\lambda/2)}$.
As in positive shifted preferential linear attachment without choice, the
asymptotic behavior of the degree distribution is non-universal, but with a
much more slowly decaying tail (Fig.~\ref{nk}).

For \emph{negative} shifted linear preferential attachment, $\lambda<0$,
(corresponding to $\frac{1}{2}<r<1$), the same analysis of the recurrence
\eqref{HD_shift} as given above predicts $n_k \sim k^{-2}$, which violates
the sum rule $\sum_{k\geq 1}kn_k\!=\!2$.  The source of this inconsistency is
that our analysis has ignored the possibility of a transition to a new type
of ``condensed'' network that contains a macrohub---a node whose degree is of
the order of $N$.  Let us assume that such a macrohub of degree $hN$ exists,
with $h$ of the order of 1.  To determine the degree of this macrohub, we now
exploit the equivalence between shifted linear attachment and the redirection
algorithm.  According to redirection, whenever a random target node is
selected, redirection will lead to the macrohub being chosen with probability
$hr$.  The probability of choosing this hub at least once in the two
independent selection events is $1-(1-hr)^2$.  This quantity gives the growth
rate of the hub, so that
\begin{equation}
\label{h2}
h = 1-(1-hr)^2\,.
\end{equation}
This equation has two solutions, $h\!=\!0$, and 
\begin{equation}
\label{hub}
h = \frac{2r-1}{r^2}~.
\end{equation}
The former (trivial) solution is relevant when the redirection probability
$r\!\leq \!\frac{1}{2}$, while the non-trivial solution \eqref{hub} is realized
when $\frac{1}{2}\!<\!r\!<\!1$.

An important feature of this macrohub is that it is unique.  To justify this
statement, suppose that more than one macrohub exists.  Denote the degrees of
the largest and second-largest hub by $h_1N$ and $h_2N$, respectively.  The
degree of the largest hub is determined from Eq.~\eqref{h2}, whose solution
is given by \eqref{hub}.  For the second-largest hub, the same reasoning that
led to Eq.~\eqref{h2} now gives
\begin{equation*}
h_2 = (1-h_1r)^2 - (1-h_1r - h_2r)^2 \,.
\end{equation*}
This equation has two solutions, $h_2=0$ and an unphysical solution
$h_2=-h_1$.  Thus a second-largest hub does not exist and greedy choice
generates one hub when $\frac{1}{2}<r<1$.

To compute the degree distribution, we must now explicitly include the effect
of the macrohub in the recurrence~\eqref{HD_shift} when $\frac{1}{2}<r<1$.
In particular, when we replace $\sum_{j<k} (j+\lambda)n_j$ by $\sum_{j\geq 1}
(j+\lambda)n_j$ as $k\to\infty$, the summation must be limited to nodes of
finite degree.  Thus we now write $\sum_{j\geq 1} (j\!+\!\lambda)n_j =
2+\lambda-h$, where the last term represents the contribution of the
macrohub.  Using the connection $\lambda=\frac{1}{r}-2$ and \eqref{hub} to
rewrite $2\!+\!\lambda\!-\!h$ as $r^{-2}\!-\!r^{-1}$, the recurrence
\eqref{HD_shift} simplifies to
\begin{equation}
\label{reduced}
n_k = -2(1-r)\,\frac{d}{dk}\,(kn_k) - 2r^2(kn_k)^2\,.
\end{equation}
The second term on the right-hand side is asymptotically negligible and the
asymptotic solution  is $n_k\sim k^{-[1+1/(2-2r)]}$.


To summarize, the degree distribution for greedy choice has the algebraic
tail
\begin{equation}
\label{n_nu}
n_k\sim k^{-\nu_2}\,,
\end{equation} 
where the decay exponent is given by (Fig.~\ref{nk}(a))
\begin{equation}
\label{nu2}
\nu_2(r)  =  
\begin{cases}
  1+1/(2r)        &  \qquad 0<r<\frac{1}{2}\,,\\
  1+1/(2-2r)     & \qquad \frac{1}{2}<r<1.
\end{cases}
\end{equation}
and the subscript refers to greedy choice from two alternatives.
Unexpectedly, $\nu_2(r)$ satisfies mirror symmetry, $\nu_2(r)=\nu_2(1-r)$.
Also notice that the two forms for $\nu_2(r)$ coincide when $r=\frac{1}{2}$.
This feature, together with the emergence of a macrohub for $r>\frac{1}{2}$
indicates that a structural transition occurs at $r=\frac{1}{2}$, and it is
natural to anticipate the appearance of a logarithmic correction at this
point, as we postulated to derive Eq.~\eqref{H_asymp}.  For comparison, in
the situation without choice, the decay exponent is $\nu_1 = 1+\frac{1}{r}$.
For the special case of strictly linear preferential attachment, $\lambda=0$
or $r=\frac{1}{2}$, the degree distribution is
\begin{equation}
n_k \simeq 4\times 
\begin{cases}
k^{-3}       &\qquad \text{no choice},\\
(k \ln k)^{-2}  &\qquad \text{binary choice}.
\end{cases}
\end{equation}

Using the above exponent $\nu_2$ in the extremal criterion \eqref{max}, the
maximal degree $k_\text{max}$ in a network of $N$ nodes with greedy choice is
given by:
\begin{equation}
\label{max2}
k_\text{max} \sim 
\begin{cases}
N^{2r}           & \qquad 0<r<\tfrac{1}{2}\,,\\
N(\ln N)^{-2} & \qquad r=\tfrac{1}{2}\,, \\
N^{2-2r}        & \qquad \tfrac{1}{2}<r<1\,.
\end{cases}
\end{equation}
The latter case actually gives the second-largest degree, as the macrohub has
the maximal degree whose value is $hN$.

To numerically implement greedy choice for shifted linear preferential
attachment, we simply allow for choice in the redirection
algorithm~\cite{GNR}.  That is, we independently identify two target nodes by
redirection and the new node attaches to the target with the higher degree.
Figure~\ref{nk}(b) shows representative simulation results for the degree
distribution with greedy choice when $r=\frac{1}{3}$ and $r=\frac{2}{3}$.
According to Eq.~\eqref{nu2}, the exponent of the two degree distributions
should be the same, as seen in our data.  For $r=\frac{2}{3}$, a unique
macrohub also emerges whose average degree is predicted from Eq.~\eqref{hub}
to be $hN$, with $h=\frac{3}{4}$.  As an illustration, simulations of 50
realizations of networks of $10^8$ nodes gives $h=0.7503\pm 0.0012$, in
excellent agreement with the theory.

\subsection{More Than Two Alternatives}

We may readily generalize to greedy choice with $p>2$ options where $p$
target nodes are selected and attachment occurs to the target with the
largest degree.  The influence of the number of options $p$ can be easily
determined for the emergence of a macrohub.  Now the analog of \eqref{h2} is
\begin{equation}
\label{hp}
h = 1-(1-hr)^p\,,
\end{equation}
from which a macrohub emerges when the redirection probability exceeds
$r_c=\frac{1}{p}$.  For $p=3$, the explicit solution is
\begin{equation}
\label{hub_3}
h=\frac{3r-\sqrt{4r-3r^2}}{2r^2}
\end{equation}
for $r>\frac{1}{3}$, while for arbitrary $p$
\begin{equation}
h\simeq \frac{2(r-r_c)}{r_c(1-r_c)}\,,  \qquad r_c = \frac{1}{p}\,.
\end{equation}
near the transition $0<r-r_c\ll 1$.  For any $p$, the macrohub degree grows
linearly in $r-r_c$ close to the transition.

For $p=3$ choices, the analog of \eqref{HD_shift} for the degree distribution
is
\begin{align}
\label{HD3_shift}
n_k &= 3\,\frac{\psi_{k-1}-\psi_k}{(2\!+\!\lambda)^3}\,\Big(\sum_{j<k} \psi_j\Big)^2 
+3\,\frac{\psi_{k-1}^2-\psi_k^2}{(2\!+\!\lambda)^3}\,\sum_{j<k} \psi_j
+\frac{\psi_{k-1}^3-\psi_k^3}{(2\!+\!\lambda)^3} + \delta_{k,1}\,,
\end{align}
with again $\psi_k=(k\!+\!\lambda)n_k$.  The first term accounts for events
where a unique maximal-degree node exists from among three choices, while the
second and third terms account for events with a two-fold and three-fold
degeneracy in the maximal-degree node, respectively.

When $-1<\lambda<1$, or equivalently $0<r<\frac{1}{3}$, the terms in the first
line of \eqref{HD3_shift} are dominant and the equation reduces to
$(kn_k)'=-\frac{1}{3}(2+\lambda)n_k$ for $k\to\infty$.  We thereby obtain
$n_k\sim k^{-[1+1/(3r)]}$.  In the marginal case of $r=\frac{1}{3}$, we again
expect a logarithmic correction of the form given in \eqref{u:def}.  With
this ansatz, the terms in the first and second lines of \eqref{HD3_shift} are
now of the same order, while the terms in the third line are negligible.  The
governing equation for $u(v)$ is
\begin{equation}
\label{DE3}
9=\Big(1-\frac{du}{dv}\Big)v^2 - 2uv\,,
\end{equation}
which gives $u = (3-v)^2 (6+v)/(3v^2)$. Combining this with
$u=\frac{dv}{d\ell}$ and specializing to the limit of large $\ell$, we find
\begin{equation}
\label{H3_asymp}
n_k \simeq \frac{3}{k^2\,(\ln k)^2}~.
\end{equation}
When $\lambda>1$ (equivalently $\frac{1}{3}<r<1$), the first term on the
right-hand side of \eqref{HD3_shift} is dominant.  However, we should again
exclude the macrohub from the sum $\Sigma_k = \sum_{j<k} (j+\lambda)n_j$.
Hence $\Sigma_k \to 2+\lambda-h$ and \eqref{HD3_shift} reduces to
\begin{equation*}
n_k = -3r[1 - hr]^2 \frac{d}{dk}\,(kn_k) \,.
\end{equation*}
Thus for the greedy three-choice model, the degree distribution scales as
$n_k\sim k^{-\nu_3}$, with
\begin{equation}
\label{nu3}
\nu_3(r)  =  
\begin{cases}
  1+1/(3r)                   &  \qquad 0<r<\frac{1}{3}\,,\\
  1+1/(3r[1 - hr]^2) & \qquad \frac{1}{3}<r<1.
\end{cases}
\end{equation}
For arbitrary $p\geq 2$, the generalization of \eqref{nu3} is
\begin{equation}
\nu_p(r)  =  
\begin{cases}
1+1/(pr)                         &   \qquad 0<r<\frac{1}{p}\,,\\
1+1/(pr[1 - hr]^{p-1})     &  \qquad \frac{1}{p}<r<1.
\end{cases}
\end{equation}
with $h\!=\!h(r)$ implicitly determined by \eqref{hp}.  In the marginal case
of $r=\frac{1}{p}$, the generalization of \eqref{H3_asymp} is
\begin{equation}
\label{Hp_asymp}
n_k \simeq \frac{p(2p-2)!}{(p-2)!}\,\, \frac{1}{k^2\,(\ln k)^2}~,
\end{equation}
and the maximal degree $k_\text{max}$ in a network of $N$ nodes is
\begin{equation}
\label{maxp}
k_\text{max} \sim 
\begin{cases}
N^{pr}                            &   \qquad  0<r<\tfrac{1}{p}\,,\\
N(\ln N)^{-2}                  &  \qquad r=\tfrac{1}{p}\,, \\
N^{pr[1 - hr]^{p-1}}        &   \qquad \tfrac{1}{p}<r<1\,.
\end{cases}
\end{equation}
As in optimization and queuing theory, the possibility of choosing between
more than two options leads only to quantitative changes compared to the more
fundamental case of two options.

\subsection{Networks With Loops}

Thus far, we studied the situation where every new node attaches to one
already existing node, leading to tree networks.  However, we can also treat
networks with loops.  Here we outline how to deal with the situation where
loops are created when each new node attaches to $m$ already existing nodes,
with each attachment event created by the same choice-driven algorithm as in
the previous section.  Limiting ourselves to shifted linear attachment and
focusing on greedy choice from two alternatives, the recursion for $n_k$ is
given by (compare with Eq.~\eqref{HD_shift})
\begin{align}
\label{nk_m}
n_k = m\,\frac{\psi_{k-1}-\psi_k}{(2m+\lambda)^2/2}
\sum_{j<k} \psi_j
-m\,\frac{\psi_{k-1}^2+\psi_k^2}{(2m+\lambda)^2}+\delta_{k,m}\,.
\end{align}

This recurrence can be analyzed using the same methods as in the case of
trees. For instance when $\lambda>0$, we replace $\sum_{j<k}
\psi_j$ by $\sum_{j\geq m} \psi_j=2m+\lambda$ when
$k\gg 1$, and then employ the continuum approximation to recast \eqref{nk_m}
into the differential equation $(kn_k)'=
-\big(1+\frac{\lambda}{2m}\big)n_k$. This equation again has an algebraic
solution of the form \eqref{n_nu}, with decay exponent
$\nu_2=2+\lambda/(2m)$.

A macrohub of degree $hN$ again emerges when $\lambda<0$, with $h$ determined
by the relation
\begin{equation}
\label{h2_m}
h = m\left[1-\left(1-\frac{h}{2m+\lambda}\right)^2\right]~,
\end{equation}
which generalizes \eqref{h2}. Thus
\begin{equation}
h = -\frac{\lambda(2m+\lambda)}{m}~.
\end{equation}
Note that the range of the shift parameter is now $\lambda>-m$, since the
minimal degree is $m$ and we must ensure that the attachment to nodes of
degree $m$ is non-negative.  The degree distribution associated with the
remaining nodes still has an algebraic tail.  To summarize, the decay
exponent is given by
\begin{equation}
\label{nu2_m}
\nu_2  =  
\begin{cases}
  2+\lambda/(2m)                             &  \qquad \lambda>0,\\
  (4m+3\lambda)/(2m+2\lambda)     & \qquad 0>\lambda> -m.
\end{cases}
\end{equation}
For the special case of strictly linear preferential attachment $\lambda=0$,
the tail of the degree distribution is
\begin{equation}
n_k \simeq 
\begin{cases}
2m(m+1)\times k^{-3}       &\qquad \text{no choice},\\
4m \times (k \ln k)^{-2}   &\qquad \text{binary choice}.
\end{cases}
\end{equation}

\section{Meek Choice}

The complementary situation of meek choice, where a set of target nodes is
first selected and a new node attaches to a target with less than the largest
degree leads to very different phenomenology.  The simplest case is that of
first selecting two nodes according to linear preferential attachment
(corresponding to $\lambda=0$) and the new node attaches to the
smaller-degree target; this specific example was also recently investigated
in~\cite{MP13}.

We determine the degree distribution in this meek choice model by following
the same approach as in greedy choice.  The analog of \eqref{HD_shift}, with
$\lambda=0$, for the degree distribution, in the case of $\lambda=0$, is
\begin{align}
\label{SD_short}
n_k = \tfrac{1}{2}\big[\psi_{k-1}-\psi_k\big]\sum_{j\geq k} \psi_j 
+\tfrac{1}{4}\big[\psi_{k-1}^2+\psi_k^2\big]+\delta_{k,1}\,.
\end{align}
Using identity $\sum_{j\geq k} jn_j = 2 - \sum_{j<k} jn_j$ recasts
\eqref{SD_short} as a recurrence.  In the case of strictly linear
preferential attachment, $\lambda=0$, the solutions for small degrees are:
\begin{align}
n_1 &= 4-2\sqrt{3}\approx 0.53589\,, \nonumber \\
n_2 &= -\tfrac{1}{2}+\sqrt{3} - \tfrac{1}{2}\sqrt{25-12\sqrt{3}}\approx 0.20548\,, \\
n_3 &=-\tfrac{1}{9}+ \tfrac{1}{3}\sqrt{25-12\sqrt{3}}
- \tfrac{2}{9}\sqrt{79-6\sqrt{25-12\sqrt{3}}   -36\sqrt{3}}\approx  0.11099\,,\nonumber
\end{align}
etc.  Notice that while the first few $n_k$ are larger than those for greedy
choice in Eqs.~\eqref{nigreedy}, the asymptotic degree distribution decays
precipitously with $k$ (Fig.~\ref{nks}).  For example, in simulations of 50
realizations of networks grown to $10^8$ nodes, the largest observed degree
is only 9!

We now exploit this rapid decay to determine the asymptotic behavior of the
degree distribution.  For large $k$, an increase in $n_k$ can occur only if
the two target nodes have degree $k-1$.  Thus we posit that the dominant term
in \eqref{SD_short} is $\tfrac{1}{4}(k-1)^2n_{k-1}^2$.  Keeping only this
term, the asymptotic behavior of the logarithm of the degree distribution is
given by
\begin{equation}
\label{exps_better}
\ln n_k \sim -C\times 2^k\,,
\end{equation}
up to some amplitude $C$ that cannot be determined within this simplified
analysis.  One can then verify that the remaining terms in \eqref{SD_short}
are subdominant.  From this asymptotic degree distribution, we estimate the
maximal degree in a network of $N$ nodes to be $k_\text{max} \simeq
\log_2\log_2 N$, as recently proven in~Ref.~\cite{MP13}.

When $p$ distinct initial target nodes are selected by preferential
attachment, there are $p$ possibilities for the attachment event: to the
highest-degree node, to the second-highest degree node, all the way to the
lowest-degree node.  While the combinatorics become unwieldy for the general
case of identifying the target node with the $m^{\rm th}$-largest degree out
of $p$ choices, the dominant contribution to $n_k$ for large $k$ arises when
$m$ targets have degree $k-1$ and the remaining $p-m$ targets have degrees
less than $k-1$.  Following the same reasoning as in the case of attaching to
the smallest-degree node out of two choices, the dominant term in the
generalization of \eqref{SD_short} is proportional to
$(k-1)^mn_{k-1}^m$. This leads to $n_k\sim \exp(-{\rm const.}\times m^k)$.
Thus for all but greedy choice, the degree distribution decays precipitously
with degree.

From this asymptotic degree distribution, the maximal degree grows with $N$
as
\begin{equation}
\label{maxp1}
k_\text{max} \sim 
\begin{cases}
N^{\omega_p}  & \text{greedy choice}\\
\log_2 \log_2 N &  2^\text{nd} ~\text{highest degree}\\
\log_3 \log_3 N & 3^\text{rd} ~\text{highest degree}\\
\cdots \\
\log_p \log_p N & \text{smallest degree}
\end{cases}
\end{equation}
for $p\geq 2$.  The exponent $\omega_p$ that appears in \eqref{maxp1} depends
on the number of alternatives $p$ and on details of the attachment rate.  For
strictly linear preferential attachment, $\omega_p = p(1-h)/(2-h)$, where the
degree $h$ of the macrohub is the positive solution of the equation $h =
1-(1-h/2)^p$.  The other ultra-slow growth laws in \eqref{maxp1} are robust
with respect to the details of the attachment rule.  These latter behaviors
do not depend on the details of the selection rule as long as the choice is
less than greedy.

\section{Summary}

Incorporating choice in preferential attachment network growth leads to rich
phenomenology in which the effect of preferential attachment can be strongly
amplified or entirely eliminated.  We have explored a general class of models
in which a set of target nodes in the network are first selected according to
preferential attachment and then a new node joins the network by attaching to
one of these target nodes according to a specified criterion.  In greedy
choice, attachment is made to the target with the largest degree.  We also
investigated attaching to a node in the target set whose degree is not the
largest.  For a target set of $p$ nodes, there are $p-1$ possible such
choices---to the $2^{\rm nd}$-largest degree node, the $3^{\rm rd}$-largest,
$\dots$, to the smallest-degree node.  We term this class of models as meek
choice.

Past work on the power of choice on the random recursive tree~\cite{RPC}
found that greedy choice broadens the degree distribution, but only in a
quantitative way.  We have shown that greedy choice plays a much more
significant role for networks that grow by preferential attachment. We
focused on shifted linear preferential attachment, but our methods apply to
other models with asymptotically linear preferential attachment. The details
depend on the model, but the general outcome is robust. In the sub-critical
phase, the degree distribution has a power law tail that is considerably
broader than in the case of no choice. In the super-critical phase, a
macrohub emerges, while the remainder of the degree distribution is still
algebraic.  At the boundary between these two phases, the degree distribution
decays as $(k\ln k)^{-2}$.  This form for the degree distribution is
consistent with a finite average degree in the network because of the
presence of the logarithmic factor.

The influence of meek choice is perhaps even more dramatic, as it effectively
counteracts preferential attachment.  When $p$ target nodes are initially
selected, meek choice means that the new node attaches to a target whose
degree is less than the highest in the target set.  For the case where a new
node attaches to the $m^{\rm th}$-largest degree out of a target set of $p$
nodes that are each selected by linear preferential attachment, meek choice
leads to a double-exponential degree distribution of the form $\exp(-{\rm
  const.}\times e^k)$, and a maximal degree that is of the order of
$\log_m\log_m N$.  It is surprising that this sharp decay should hold for
attachment to the target with the $2^{\rm nd}$-highest degree out of $p\gg1$
targets.  In this case, the degree distribution will initially resemble that
of greedy choice and the crossover to a precipitous decay will occur at an
extremely large degree value.

\bigskip\noindent This research was partially supported by the AFOSR and
DARPA under grant \#FA9550-12-1-0391 and by NSF grant No.\ DMR-1205797.

\end{document}